\documentclass[sigconf, nonacm]{acmart}
\usepackage{graphicx}
\usepackage{textcomp}
\usepackage{xcolor}
\usepackage{subfigure}

\usepackage[ruled,vlined,linesnumbered]{algorithm2e}

\SetKwFunction{FExecLTX}{EXEC\_LTX}
\SetKwFunction{FExecOCC}{EXEC\_OCC}
\SetKwFunction{FExecRTX}{EXEC\_RTX}
\SetKwFunction{FValidatePhantom}{validate\_phantom\_freedom}
\SetKwInput{Input}{Input}
\SetKwInput{Output}{Output}
\SetKw{And}{and}
\SetKw{Or}{or}

\makeatletter
\newcommand{\linebreakand}{%
  \end{@IEEEauthorhalign}
  \hfill\mbox{ }\par\vspace{1em}
  \mbox{ }\hfill\begin{@IEEEauthorhalign}
}
\makeatother

\begin{document}

\title{Shirakami: A Hybrid Concurrency Control Protocol for \\Tsurugi Relational Database System}

\author{Takayuki Tanabe*}\thanks{*Work done while at NAUTILUS Technologies, Inc}
\affiliation{%
  \institution{Keio University}
}
\email{tanab@keio.jp}

\author{Shinichi Umegane}
\affiliation{%
  \institution{NAUTILUS Technologies, Inc}
}
\email{umegane@nautilus-technologies.com}

\author{Suguru Arakawa}
\affiliation{%
  \institution{NAUTILUS Technologies, Inc}
}
\email{arakawa@nautilus-technologies.com}

\author{Ryoji Kurosawa}
\affiliation{%
  \institution{NAUTILUS Technologies, Inc}
}
\email{kurosawa@nautilus-technologies.com}

\author{Takashi Hoshino}
\affiliation{%
  \institution{Cybozu Labs, Inc}
}
\email{hoshino@labs.cybozu.co.jp}

\author{Hideyuki Kawashima}
\affiliation{%
  \institution{Keio University}
}
\email{river@sfc.keio.ac.jp}

\author{Masahiro Tanaka}
\affiliation{%
  \institution{Keio University}
}
\email{masa16.tanaka@keio.jp}

\author{Takashi Kambayashi}
\affiliation{%
  \institution{NAUTILUS Technologies, Inc}
}
\email{kambayashi@nautilus-technologies.com}

\begin{abstract}
Bill-of-materials and telecommunications billing applications need to process both short and long read-write transactions simultaneously.
Recent work rarely addresses such evolving workloads. 
To deal with these workloads, we propose a new concurrency control protocol, Shirakami. Shirakami is a hybrid protocol. The first protocol, Shirakami-LTX, is for long read-write transactions based on multiversion view serializability. The second protocol, Shirakami-OCC, is for short transactions based on Silo. Shirakami naturally integrates them with write-preservation and epoch-based synchronization. It does not require dynamic protocol switching and provides stable performance.
We implemented Shirakami as the transaction processing module of the Tsurugi system, which is a production-grade relational database system.
The experimental results demonstrated that Tsurugi was at least 7.9$\times$ faster than PostgreSQL, and Shirakami-LTX exhibited 680$\times$ higher throughput than Shirakami-OCC.
\end{abstract}
\maketitle

\section{Introduction}
\subsection{Motivation}

\textcolor{black}{
A new class of transactional workloads has emerged: the combination of long read-write transactions and short transactions.
Currently, long read-write transactions are typically isolated from concurrent online transaction processing (OLTP) traffic by operational means, such as nightly batch windows or stale materialized views, rather than by the concurrency control protocol itself. 
As real-time and on-demand processing becomes standard, these workarounds are increasingly impractical. 
}

\textcolor{black}{
The algorithmic alternatives have limits. Snapshot Isolation, adopted in commercial engines, admits write skew and cannot guarantee consistency in the bill of materials workload, which requires long read-write transactions~\cite{oze-vldb}. 
Conversely, in-memory engines~\cite{hekaton} and classical protocols~\cite{lim2017cicada, wang2017efficiently, Chen22plor, epic, tianzheng25} are optimized for short transactions and exhibit excessive aborts on long ones~\cite{oze-vldb}. Neither family alone supports the long-short coexistence we target.
}
Modern protocols such as TuskFlow~\cite{tuskflow} and \textcolor{black}{Data-Driven-Isolation (DDI)~\cite{ddi}} present high performance in long read-write transactions, but they can be applied only to graph databases.
Oze~\cite{oze-vldb} is for general cases, but it has not been implemented in a production system.
What this paper addresses is how to make \textbf{general-purpose}, \textbf{long read-write} transactions coexist with short transactions while maintaining consistency within the \textbf{production} database.

This paper addresses two cases. The first case is the Bill of Materials (BoM) for the bread manufacturing industry~\cite{oze-vldb}. It is based on an item master and an item configuration master.
In BoM, cost calculation is performed for all registered items.
BoM needs to read many records and write the results; thus, the update process takes a long time to execute.
In addition to these long-term transactions, BoM performs short-term transactions to update the raw material costs referenced in the calculation of product costs, as well as short-term transactions to update product costs in other applications.
The second case is cell phone billing processing in a telecommunications company.
When calculating call charges, the company matches the call history, which is updated in real time, with customer contract information to determine which subscriptions are to be billed so that the call history can be billed accurately.
It is also not possible to determine in advance which data needs to be read for this process.
%

\subsection{Problem}
Current general-purpose protocols do not address these workloads. As shown in \cite{oze-vldb}, classical protocols~\cite{tu2013speedy, kim2016ermia, yu2016tictoc, lim2017cicada} do not process long-term transactions efficiently because their transaction scheduling spaces are limited to conflict serializability or a multiversion space narrower than multiversion view serializability. To our knowledge, only the Oze protocol~\cite{oze-vldb} performs long-term read-write transactions due to its wide scheduling space, exploiting the \textcolor{black}{multi-version serialization graph (MVSG)~\cite{Bernstein}}. 
However, Oze does not support the concurrent operation of long- and short-transaction protocols at the same time. Under a mixed workload, every transaction must be processed by its long-transaction protocol, which is inefficient for short transactions and degrades latency.

Some work on graph data uses snapshot isolation for long read-only transactions. However, it does not work for long read-write transactions over large-scale graphs. Recent work addresses this based on a deterministic protocol~\cite{tuskflow} or flexible isolation-level management~\cite{ddi}. These approaches are designed for graph management and are not intended for general-purpose use.

Modern cloud-native databases \textcolor{black}{HiEngine~\cite{hiengine}} and \textcolor{black}{Redshift~\cite{redshift-reinvented}} are designed based on Silo~\cite{tu2013speedy} and ERMIA~\cite{kim2016ermia}, respectively. Therefore, both are almost incapable of successfully handling long batch transactions. Consequently, companies are effectively forced to execute huge transactions, called batches, at night, separately from daily data processing.
What kind of protocols should be designed to handle real-world workloads that include both long read-write transactions and short transactions? The purpose of this study is to answer this question.

\subsection{Approach and Contribution}
To enable large batch transactions and relatively short, intermittent, and continuous online transactions to coexist, we propose integrating separate protocols for each. We refer to this hybrid protocol as the Shirakami protocol.
It has a protocol for long transactions, \textcolor{black}{Shirakami-LTX (S-LTX)}, and a protocol for short transactions, \textcolor{black}{Shirakami-OCC (S-OCC)}.

S-LTX can handle long transactions such as BoM~\cite{oze-vldb}. The scheduling space of many modern protocols is conflict serializability (CSR); in this paper, we use a larger space than the one generated by the \textcolor{black}{multi-version timestamp ordering (MVTO)} concurrency control protocol.
Decisions of serializability based on CSR contain a huge number of false positives, while \textcolor{black}{multi-version view serializability (MVSR)} has no false positives. To our knowledge, only theoretical multiversion serial graph methods~\cite{weikum2001transactional} and Oze~\cite{oze-vldb} use this wide space, which is larger than the space generated by MVTO, and our proposal is the first production-level implementation of this space.

S-LTX performs well for long transactions by utilizing the large scheduling space, but it does not exhibit the same efficiency for short transactions as modern CSR-based protocols.
To overcome this weakness of S-LTX, we introduce another protocol, S-OCC, an efficient protocol for short-term transactions such as YCSB or TPC-C. S-OCC applies an optimistic technique like Silo with modifications that allow co-existence with S-LTX.

S-LTX and S-OCC can run concurrently. Therefore, we can apply S-OCC for short transactions and S-LTX for batch transactions concurrently and efficiently.
Our scheme enables efficient transactions across any workload.
%
We first evaluated Tsurugi, and it was \textcolor{black}{5.0}$\times$ and at least \textcolor{black}{7.9}$\times$ faster than PostgreSQL in the Bill of Materials Benchmark~\cite{oze-vldb} and the Phone Billing Benchmark, respectively.
We also evaluated the performance of S-LTX compared to S-OCC with an experiment based on YCSB with long transactions.
The experimental results with Shirakami demonstrated that S-LTX exhibited 680$\times$ higher throughput than S-OCC.
The whole Tsurugi system, including Shirakami, is available online~\cite{tsurugi_git}.

\subsection{Organization}
The rest of this paper is structured as follows.
Section \ref{sec:prep} describes the preliminaries.
Section \ref{sec:proposal} describes the proposed method.
Section \ref{sec:Evaluation with PostgreSQL} evaluates Tsurugi.
Section \ref{sec:eval} evaluates Shirakami.
Section \ref{sec:related} describes related work.
Section \ref{sec:concl} concludes this paper.

\section{Preliminaries}
\label{sec:prep}
\subsection{Motivating Use Cases: Long Transactions}
\label{sec:Motivating Use Cases: Long Transactions}
Transactions that perform analytical processing are executed in our motivating applications (\textcolor{black}{bill of materials~\cite{oze-vldb}} and phone billing in $\S$\ref{sec:Evaluation with PostgreSQL}). 
They perform large-scale read and write operations.
In business systems such as inventory control, order receipt or placement, customer management, human resource payroll, logistics, and financial transactions, a variety of tasks are executed.
They include cut-off processing, valuation processing, cost management, salary slip calculation, order summary, invoicing, etc.

\textcolor{black}{
Serializable execution is mandatory in transaction processing to avoid concurrency anomalies, while high performance often requires concurrent execution of multiple transactions. Some workloads include heavy transactions that scan numerous records across multiple large tables, compute summaries, and write the results back. These workloads are therefore characterized by the coexistence of short transactions and large read-write transactions. 
}

We address two cases as long update transactions.
One is a batch-costing approach for the food processing industry using BoM~\cite{oze-vldb}, which traverses tree-structured data and accumulates product costs.
The other calculates cell phone charges using a telecom company's billing process. It is a mixed workload of batch and online processing that simultaneously calculates and aggregates customer usage charges for all customers while phone usage transactions are continuously recorded.
This processing is usually done separately from online processing at night because updates from online processing conflict during batch processing. The mixed workload must support real-time aggregation of rate calculations.

\subsection{Concurrency Controls}
\subsubsection{Protocols for Short Transactions}
Modern \textcolor{black}{optimistic concurrency control (OCC)} is known to exhibit excellent performance when the number of operations in a transaction is small~\cite{lim2017cicada, Chen22plor, Masumura22}.
The features of modern OCC, which originated with Silo~\cite{tu2013speedy}, include avoiding deadlocks by aligning the lock order immediately before the lock phase, eliminating or mitigating the use of centralized counters, adopting invisible reads~\cite{CCBench}, and shortening the critical section by early lock release~\cite{Nakamori2023}.

\subsubsection{Protocols for Long Transactions}

\textcolor{black}{
Long-term transactions suffer from increased conflict probability. While techniques such as Cicada~\cite{lim2017cicada} and Serial Safety Net (SSN)~\cite{wang2015serial} offer broader scheduling spaces than CSR, they remain inefficient for long transactions such as BoM. This stems from scheduling spaces that are still too narrow to accommodate long transactions, as discussed in $\S$2.6 in \cite{oze-vldb}.
}

An interesting approach is the deterministic protocol, which assumes that the entire contents of a transaction are known before submitting it to the database~\cite{Thomson12calvin, Lu2020Aria, tuskflow}.
The execution order within a set of transactions is determined before they are submitted to the system, and the transactions are executed serially. Only when there are no conflicts can multiple transactions be executed in parallel.
This method is efficient if the data accessed by the transactions is invariant, but in reality, this may not be the case. In such cases, the performance of the deterministic approach deteriorates significantly~\cite{oze-vldb}.

\section{Proposal: Shirakami}
\label{sec:proposal}
\subsection{Overview}

To address our workload, which consists of both short transactions and long read-write transactions, we present Shirakami, a hybrid concurrency control protocol.
The first protocol is \textcolor{black}{Shirakami-OCC (S-OCC)}, which is a modification of 
\textcolor{black}{Silo~\cite{tu2013speedy}.
Silo is SERIALIZABLE, and so is S-OCC.
} 
The second protocol is \textcolor{black}{Shirakami-LTX (S-LTX)}, a newly designed protocol for handling long read-write transactions.
We describe them as \textbf{S-OCC} and \textbf{S-LTX}, respectively.
The Shirakami concurrency control protocol can commit long transactions that are difficult to commit in a contended environment with conventional protocols, while the performance of processing short transactions is not degraded compared to those protocols.

\subsubsection{Priority}
When S-OCC and S-LTX transactions run concurrently, they are scheduled according to their priorities.
There are two priority levels, and S-LTX has a higher priority than S-OCC.
%
\textcolor{black}{
%
This is because Shirakami primarily targets workloads consisting of a single long transaction and multiple short transactions. These workloads include BoM and Billing, as shown in the paper. In the presence of numerous S-LTX, S-OCC will constantly compete with S-LTX and may become starved. 
To avoid this, applications can execute starved S-OCC transactions as new S-LTX transactions or limit the flow rate of S-LTX.
}

%
%
\textcolor{black}{
For S-LTX transactions, the one that starts earlier has a higher commit priority. 
Suppose there are two S-LTX transactions with overlapping lifetimes, and it is impossible to commit both of them. In that case, the S-LTX transaction that takes priority commits first. 
The other is aborted.
}
%

\subsubsection{Staging S-LTX}
After submitting S-LTX transactions, they do not participate in the ongoing epoch; instead, they start at the beginning of the next epoch.
They have higher priority than all running and future S-OCC transactions.
The S-LTX transactions to be executed at that time share the snapshot to read, which is the result of the previous epoch. Shirakami uses a multiversion data structure to provide snapshots per epoch.

All S-LTX transactions declare the area to be written at their starting time. We refer to this as \textbf{write preservation or WP}.
If a write operation of an S-LTX transaction conflicts with a read of an S-OCC transaction, the S-OCC transaction can detect all writes of the S-LTX transaction before reading them. Thus, as the original Silo waits for the locked object in the read phase,
S-OCC transactions can abort when they find WP information indicating a conflict in the same epoch.

\subsubsection{Order Forwarding}
\label{sec:Order Forwarding}
\begin{figure}[t]
 \centering
 \includegraphics[scale=0.090]{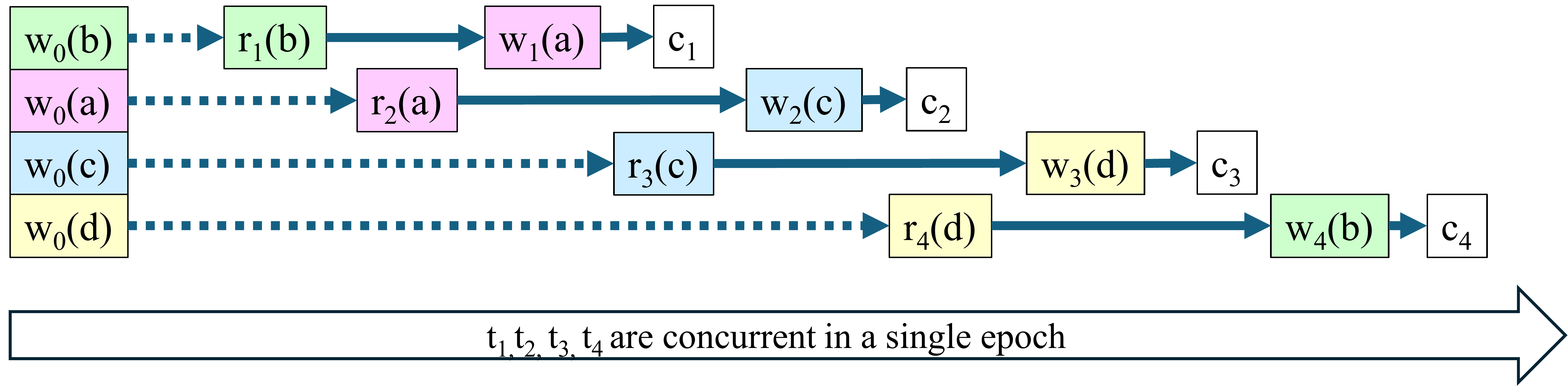}

\caption{\small An example of false positive aborts.
  Four transactions ($\mathrm{t}_1,\ldots,\mathrm{t}_4$) are running concurrently. $\mathrm{w}_0(x)$~\cite{weikum2001transactional} expresses the initial value of record $\mathrm{x}$. 
  $\mathrm{c}_\mathrm{k}$ denotes partial commit (finalized in the next epoch).
  With S-LTX, $\mathrm{t}_1$, $\mathrm{t}_2$, and $\mathrm{t}_3$ commit; despite committing in temporal order $\mathrm{t}_1$, $\mathrm{t}_2$, $\mathrm{t}_3$, their serialization order becomes $\mathrm{t}_3 \to \mathrm{t}_2 \to \mathrm{t}_1$
  via order forwarding. $\mathrm{t}_4$ aborts since $\mathrm{r}_4(\mathrm{d})$ prevents placing $\mathrm{t}_4$ before $\mathrm{w}_0(\mathrm{b})$. 
  With S-OCC, $\mathrm{t}_1$ and $\mathrm{t}_3$ commit; $\mathrm{t}_2$ and $\mathrm{t}_4$ abort due to
  conflicts $(\mathrm{r}_2(\mathrm{a}), \mathrm{w}_1(\mathrm{a}))$ and $(\mathrm{r}_4(\mathrm{d}), \mathrm{w}_3(\mathrm{d}))$, respectively. 
  With MVTO, only $\mathrm{t}_4$ commits because $\mathrm{r}_2(\mathrm{a})$, $\mathrm{r}_3(\mathrm{c})$, and $\mathrm{r}_4(\mathrm{d})$ prevent $\mathrm{w}_1(\mathrm{a})$, $\mathrm{w}_2(\mathrm{c})$, and $\mathrm{w}_3(\mathrm{d})$, respectively.
}

 \label{fig:Running Example of Shirakami}
\end{figure}

Concurrent S-LTX transactions can conflict when they access the same record, specifically when a higher-priority transaction ($t_1$) writes it while a lower-priority transaction ($t_2$) reads it (leading to different versions). In this case, $t_2$ is placed before $t_1$ in the serialization order, and $t_2$'s serialization epoch is adjusted to match $t_1$'s serialization epoch.
We refer to this as \textit{order forwarding}.
This makes it possible for $t_2$ to commit without being aborted by $t_1$'s write, even though $t_1$ has a higher commit priority.
This order forwarding enables achieving a serialization order outside the CSR scheduling space.

Figure \ref{fig:Running Example of Shirakami} 
shows examples of false-positive aborts.
Here, four transactions are running concurrently. All transactions occur within a single epoch.
The different colors indicate different records.
\textcolor{black}{
In the figure, consider which transactions can succeed under three protocols: S-LTX, S-OCC, and MVTO.
}
Using S-LTX, three transactions are committed, which is ideal and incurs no false positives. This is due to the wide scheduling space provided by order forwarding.
Using S-OCC or Silo~\cite{tu2013speedy}, we can commit only two transactions and incur one false positive.
Using MVTO~\cite{lim2017cicada,weikum2001transactional}, we can commit only one transaction.
There are two false positives in this case.

\subsubsection{Details of Write Preservation (WP)}
Conflict avoidance is essential for long-term transactions.
\textcolor{black}{Two-phase locking (2PL)}-like locking ensures conflict avoidance but may increase aborts and retries for the sake of deadlock avoidance.
OCC can exhibit excellent performance for short transactions, but it results in more aborts because it cannot detect conflicts until the commit phase.
A deterministic approach can reduce conflicts and is thus well-suited for long transactions; however, its performance on short transactions tends to be limited.

An intermediate technique is to preserve write operations by placing write clues that are similar to the write quorum~\cite{Fan19oceanvista, tapir}.
\textcolor{black}{
A short OCC transaction aborts itself when it detects a conflicting write preservation clue from a higher-priority long transaction, because short transactions are cheap to retry.
}
If the write set of a transaction is predetermined, the transaction leaves only its clue on the target records.
Using this as a hint, the read operation of short transactions can observe future write operations by long transactions.
%
%
All subsequent transactions can refer to that declaration to avoid invalidating prior transactions due to conflicts.


\textcolor{black}{
\subsubsection{Granularity of WP: Table Level}
We adopt table-level granularity for write preservation. 
The false positives that occur in S-OCC's read validation due to the table-level granularity of S-LTX's WPs are unavoidable under Shirakami's chosen design (1)(2).
(1) When S-LTX and S-OCC operate concurrently, a typical S-OCC would want to commit first, but WPs are necessary to avoid interfering with the (future) commits of the S-LTX, which should have higher priority.
(2) The assumption that the target to be accessed at the record level can be identified at the start of S-LTX cannot handle situations where the write set is not determined, so a design decision was made to handle WPs at the table level.
A comparison with a design that uses WPs at the record level only for predictable workloads will be in future work, and a quantitative discussion of false positives will be necessary at that time.
While false positives of S-OCC transactions against S-LTX transactions are unavoidable, S-LTX transactions themselves are validated at a finer granularity rather than at the table level.
The design prevents commit-order reversals between S-LTX transactions, except for read-area declarations, allowing applications to observe the write set of committed S-LTX transactions during validation.
If aborting specific S-OCC transactions becomes a problem, applications can avoid the S-OCC false positive issue by executing them as S-LTX transactions, albeit at the expense of overhead.
}

\textcolor{black}{
\subsubsection{Transaction Mode Selection}
Shirakami currently does not provide runtime mode inference; instead, the application developer designates each transaction as either S-LTX or S-OCC at submission time. We recommend the following heuristic for this designation. 
Transactions that are known a priori to read or write a large number of records — typical examples are nightly batch jobs, the cost-accumulation of BoM ($\S$4.3), and the phone billing batch of telecommunication ($\S$4.2) — should be submitted as S-LTX. Transactions that touch only a handful of records and are expected to be issued at high frequency — typical examples are
OLTP-style point queries and updates, such as those in
TPC-C ($\S$4.4) — should be submitted as S-OCC. The Shirakami evaluation in $\S$5.2 provides empirical support for this heuristic.
}

\textcolor{black}{
A mis-designation does not compromise correctness, since both S-LTX and S-OCC are SERIALIZABLE ($\S$3.7); only performance is affected. Should a designation turn out to be suboptimal at runtime, applications can take the following actions without changing an already-running transaction's mode. (i) An S-OCC transaction that has been starved by competing S-LTX transactions, or whose access pattern is found to conflict with S-LTX too frequently ($\S$3.3.3 Case 3), can be retried as a fresh S-LTX transaction ($\S$3.7). (ii) The application can control the volume and ordering of S-LTX submissions — limiting the rate protects concurrent S-OCC transactions from starvation, and submitting more important S-LTX transactions earlier exploits their earlier-first commit priority ($\S$3.1.1, $\S$3.7).
}

%
%
%
%
\subsection{Shirakami-LTX Protocol}
\label{sec:Shirakami-LTX}
The Shirakami-LTX (S-LTX) is designed for efficiently processing long read-write transactions. It has an execution phase, an order-forwarding phase, and a write-validation phase, as shown in Algorithm \ref{alg:shirakami-ltx}.
A notable feature of S-LTX is order forwarding based on priority management.
When two S-LTX transactions are running, and a low-priority transaction tries to read data written by a high-priority transaction,
the low-priority transaction attempts order forwarding (i.e., a dynamic change of serialization order).
%
If order forwarding succeeds, its serialization epoch is set to the high-priority transaction's epoch, and the two transactions are ordered by reverse priority within that epoch.

\subsubsection{Execution Phase}
The execution phase is described in lines \ref{ln:exec-s} to \ref{ln:exec-e} of Algorithm \ref{alg:shirakami-ltx}.
When starting a transaction, S-LTX temporarily pauses the epoch's progress with a lock.
It obtains the current epoch $N$ and sets its own startable epoch (initial value of the serialization epoch, valid epoch) to $N + 1$. Then it registers its own transaction ID and the startable epoch as WP in the table-level metadata, and finally releases the lock on the epoch.

The core of this design is to suppress global epoch advancement to prevent exposure of intermediate states during incomplete WP registration. If the global epoch advances while a transaction is still registering its WP, a race condition could arise: a subsequent OCC read might bypass validation by failing to detect the unregistered WP and commit prematurely, only for the WP registration to complete immediately thereafter. 

For reading data, it first checks the table-level WP and, if it finds a WP with a higher priority than its own, reserves its transaction ID for S-LTX in the local set. It also adds its own reads to the full-scan, range-scan, or point-search information as the read range. This information is used to eliminate the phantom reads anomaly. The details are described in $\S$\ref{sec:Phantom Avoidance}.
For writing data, it saves them in the local write set.

\begin{algorithm}[tb]
\caption{Shirakami-LTX Protocol (EXEC\_LTX)}
\label{alg:shirakami-ltx}
\DontPrintSemicolon
\small
\SetInd{0.3em}{0.7em}
\Input{A transaction $tx$ with $tx.type = \text{LTX}$}
\Output{COMMIT or ABORT}
$\text{lock\_epoch\_advance}()$\nllabel{ln:begin-s};
$tx\_id \gets \text{assign\_transaction\_id}()$
    \tcp*{\S3.2.1}
$commit\_epoch \gets \text{get\_global\_epoch}() + 1$;
$\text{register\_wp}(tx.wp\_set, tx\_id, commit\_epoch)$\;
$\text{unlock\_epoch\_advance}()$;
\textbf{wait until} $\text{get\_global\_epoch}() \ge commit\_epoch$\nllabel{ln:begin-e}
    \tcp*{staging (\S3.1.2)}
$overtaken\_set, read\_set, write\_set \gets \emptyset$;
$max\_read\_epoch \gets 0$;
$forwarded \gets \text{false}$\nllabel{ln:exec-s}\;
\ForEach(\tcp*[f]{execution phase}){$opr \in tx$}{
    \eIf{$opr.type = \text{READ}$}{
        $overtaken\_set \gets overtaken\_set \cup \{id \in \text{WP}(opr.key) : id < tx\_id\}$\;
        $value, ver \gets \text{read\_snapshot}(opr.key, commit\_epoch)$\;
        $read\_set \gets read\_set \cup \{opr.key\}$;
        $max\_read\_epoch \gets \max(max\_read\_epoch, ver.epoch)$\;
    }{
        $write\_set \gets write\_set \cup \{opr\}$\nllabel{ln:exec-e}\;
    }
}
$\text{wait\_for\_completion}(overtaken\_set)$\nllabel{ln:of-s}
    \tcp*{wait for higher-priority LTXs}
\ForEach(\tcp*[f]{order forwarding}){$ltx\_id \in overtaken\_set$}{
    \If{$\text{committed}(ltx\_id) \land (read\_set \cap \text{write\_keys}(ltx\_id) \ne \emptyset)$}{
        $forwarded \gets \text{true}$;
        $new\_epoch \gets \min(commit\_epoch, \text{commit\_epoch\_of}(ltx\_id))$\;
        \If{$max\_read\_epoch \ge new\_epoch$}{ \Return{ABORT} \tcp*{read validation} }
        $commit\_epoch \gets new\_epoch$\nllabel{ln:of-e}\;
    }
}
\If(\tcp*[f]{write validation: only if forwarded (\S3.7)}){$forwarded$\nllabel{ln:wv-s}}{
    \ForEach{$opr \in write\_set$}{
        \If{$commit\_epoch \le \text{max\_reader\_epoch}(opr.key)$}{ \Return{ABORT}\nllabel{ln:wv-e} }
    }
}
$\text{register\_reader\_epoch}(read\_set, commit\_epoch)$;
$\text{apply\_writes}(write\_set, commit\_epoch)$\;
$\text{publish\_wp\_result\_and\_remove\_wp}(tx\_id, write\_set)$\;
\Return{COMMIT}\;
\end{algorithm}

\subsubsection{Commit Protocol: Order Forwarding and Write Validation}
The commit phase verifies whether a transaction can be committed. It first conducts the order forwarding phase (lines \ref{ln:of-s}--\ref{ln:of-e}).
An S-LTX transaction first invokes order forwarding (as described in $\S$\ref{sec:Order Forwarding}) if required.
\textcolor{black}{
S-LTX transactions are verified in priority order
to prevent a scenario where a low-priority transaction commits before a high-priority one, subsequently causing the latter to abort.
}
%

The wait processing is conducted as follows.
If there is a transaction with a higher priority than the transaction that writes or places a WP, and the lower limit of order forwarding for the higher-priority transaction is not fixed, then the lower limit of order forwarding for the current transaction is not fixed either.
Thus, validation is delayed until the lower limit of order forwarding of the transaction in question is fixed.

In read validation, if another transaction has already written a new version or placed a WP, the transaction attempts to order-forward.
If the transaction exceeds its lower limit for order forwarding, it aborts.
%
%
The lower limit of order forwarding is the minimum value among all transactions that overwrite or place WPs on the current transaction's read set.
%
%
Finally, write validation is executed only when the transaction has performed order forwarding (lines \ref{ln:wv-s}--\ref{ln:wv-e}); the justification for skipping it otherwise is given in §3.7.

\subsubsection{Phantom Avoidance}
\label{sec:Phantom Avoidance}
S-LTX checks for phantoms from the writer's side but not from the predicate reader's side. Predicate reads are managed in three types: full scans, range scans, and point searches. They can be referenced by registering the type of predicate read in each table at the time of reading.

The writer's side requires a mechanism to avoid phantoms. Consider the case where the S-LTX transaction $t_1$ with a predicate read has been committed, and another S-LTX transaction $t_2$ is in the write verification phase. $t_1$ has a higher priority than $t_2$. If $t_2$ precedes $t_1$ in serialization order by order forwarding, $t_1$'s predicate read observes a phantom after the fact. Therefore, $t_2$ aborts itself.

The predicate reader side does not require an additional mechanism for phantom avoidance. When a low-priority S-LTX transaction performs a predicate read on a record written by a high-priority S-LTX transaction, the record is resolved by WP. As for concurrent S-OCC transactions, uncommitted S-OCCs at the start of an S-LTX are guaranteed to be placed after the S-LTX in serialization order, so S-OCC writes never affect S-LTX reads.

%
%
%
%

\subsection{Shirakami-OCC Protocol}
For reads in a successful Shirakami-OCC (S-OCC) transaction execution, as long as there is no S-LTX transaction that conflicts with the S-OCC transaction at commit time (in the case of a conflict, the S-OCC transaction is guaranteed to lose), there is no problem since the order is descending in the serialization order.
%
%
%
However, an S-LTX transaction may be placed earlier than its opening epoch by order forwarding. When an S-OCC transaction has already committed but is subsequently placed after the S-LTX transaction in serialization order, the S-LTX transaction must check the S-OCC transaction's read set during write validation.

For this purpose, S-OCC transactions store metadata indicating which data they have read, even after committing, for S-LTX write validation. 
\textcolor{black}{
Details are described in Algorithm \ref{alg:shirakami-occ}.
Since this metadata is updated in place, garbage collection is not required. For instance, the metadata only maintains the maximum value for read timestamps.
}

\subsubsection{Execution Phase}
The execution phase is described in lines 2 to 8 of Algorithm \ref{alg:shirakami-occ}.
It is similar to the read phase of Silo, except for read-access management.

Please note that a straightforward approach to storing information about which data items are read by transactions incurs resource waste.
It increases memory usage and reduces cache utilization.
%

To address this issue, we set each record to hold only the latest read epoch information (std::atomic$<$epoch$>$), freeing it from garbage collection and preventing cache pollution within the same epoch. This approach allows Silo to store read information with a worst-case performance degradation of about 5\%.

\subsubsection{Commit Phase}
The commit phase of S-OCC (lines 9--17) is basically the same as that of Silo~\cite{tu2013speedy}, except that 
S-OCC checks for write preservation in read regions and registers the reader epoch.
%
The check of the WP \textcolor{black}{guarantees} that the write preservation to be observed is always globally verified.
In addition, the lock-free WP design allows WP information to be loaded lock-free and atomically. This validation is not expensive since the WP is at the table level.

\subsubsection{Phantom Avoidance}
\label{sec:Phantom Avoidance OCC}

Since an S-LTX transaction performs order forwarding, it can cause phantom reads for S-OCC transactions. To prevent this, S-OCC must validate its predicate reads during the commit phase to ensure that no S-LTX transaction has introduced phantoms.

\noindent
\textbf{Case 1: S-LTX is still live (before commit).}
\quad
If an S-LTX transaction has not yet committed when an S-OCC transaction's commit validation occurs, the S-LTX's WP remains present in the table-level metadata. Since WP is a per-table declaration, any predicate read by S-OCC will observe it, making the conflict detectable via the WP check.

\noindent
\textbf{Case 2: S-LTX has already committed (WP erased).}
\quad
If an S-LTX transaction has already committed and erased its WP metadata, the individual node updates have already been applied. Therefore, the conflict can be detected via node-set validation using the Silo phantom-avoidance mechanism in the S-OCC commit protocol. Since WP erasure occurs only after node updates are complete, node verification will always detect the conflict.

\noindent
\textbf{Case 3: S-OCC has already committed.}
\quad
An S-LTX transaction that tries order forwarding checks the table-level read clues stored by an S-OCC transaction that has performed predicate reads. If the forwarding breaks the S-OCC transaction's predicate reads, the S-LTX transaction aborts.
\textcolor{black}{
This overhead is minimal: each table maintains a single read clue updated at most once per epoch, requiring only a single read by S-LTX. Such occurrences are rare, as they require specific "order forwarding" scenarios involving complex multi-transaction conflicts. If this pattern becomes frequent, it can be easily mitigated by promoting the S-OCC range scans to S-LTX.
%
}

\begin{algorithm}[tb]
\caption{Shirakami-OCC Protocol}
\label{alg:shirakami-occ}
\DontPrintSemicolon
\small
\SetKwBlock{FnOCC}{function \FExecOCC{$tx$}}{end}
\FnOCC{
    \tcp{Execution Phase}
    $read\_set \gets \emptyset, write\_set \gets \emptyset$\;
    \ForEach{$opr \in tx$}{
        \eIf{$opr.type = \text{READ}$}{
            $value, version \gets \text{read\_latest}(opr.key)$\;
            $read\_set \gets read\_set \cup \{\langle opr.key, version \rangle\}$\;
        }{
            $write\_set \gets write\_set \cup \{opr\}$\;
        }
    }

    \tcp{Commit Phase}
    $\text{acquire\_write\_locks}(write\_set)$ \tcp*{Acquired in key order}
    $commit\_epoch \gets \text{get\_global\_epoch}()$\;

    \ForEach{$\langle key, version \rangle \in read\_set$}{
        \If{$version \ne \text{current\_version}(key) \lor \text{wp\_conflict}(key, commit\_epoch)$}{
            $\text{release\_write\_locks}()$\;
            \Return{ABORT}\;
        }
    }

    $\text{apply\_writes}(write\_set)$\;
    $\text{register\_reader\_epoch}(read\_set, commit\_epoch)$\;
    $\text{release\_write\_locks}()$\;
    \Return{COMMIT}\;
}
\end{algorithm}

\subsection{Epoch}
The Shirakami protocol manages epochs as fixed time units to handle groups of transactions. A transaction has a serialization point in an epoch.
\textcolor{black}{This is also used in Silo~\cite{tu2013speedy}.}
Within each epoch, the serialization order of a group of transactions is determined, and the version that should be persisted at the end of the epoch is established.
In each epoch, records to be persisted are passed to the logger, which asynchronously synchronizes them for efficiency.
Additionally, the logging system creates the necessary snapshots.

\textcolor{black}{
\subsubsection{Epoch Duration}
The default epoch setting is 3 ms. The epoch length can be adjusted, but there are trade-offs when doing so. Setting a longer epoch reduces epoch carryover and the cost of garbage collection in multi-version environments. On the other hand, it limits S-LTX's execution throughput. This is because even after an S-LTX finishes, subsequent S-LTX executions will be in a waiting state until the current epoch is complete. Setting a shorter epoch reverses these advantages and disadvantages.
}

\subsubsection{Logging}
At the end of an epoch, a set of pre-committed records is passed to the logging system \textit{Limestone},
for asynchronous persistence. Since our implementation uses the write-ahead logging protocol, pre-write is performed. However, to alleviate unnecessary version pressure, the non-visible write rule~\cite{nakazono2019invisiblewriterule} is used.

\subsubsection{Snapshot}
An S-LTX transaction is assigned to a specific epoch at the start and reads the specific version of each record for that epoch. Unlike persisted snapshots (used for fault tolerance and availability), this is an in-memory snapshot: a set of committed versions maintained solely for transaction reads.

Shirakami uses two types of in-memory snapshots.
The first type is the unsafe snapshot.
In-memory snapshots are created at every epoch and consist of the latest committed version at the end of that epoch.
The created snapshot may be inconsistent because S-LTX transactions that run later and are moved to the epoch via order forwarding may change it.
If a transaction reads a record in an inconsistent snapshot, the transaction is aborted.
S-LTX transactions currently read unsafe snapshots, taking the abort risk but preferring newer snapshots.

The second type is the safe snapshot.
If no S-LTX transactions execute order forwarding at the end of an epoch, a snapshot consisting of the latest committed versions is referred to as a safe snapshot.
Record sets in safe snapshots can be read safely, so their read validation is not aborted.
When the application designates a transaction as read-only, Tsurugi executes it against a safe snapshot, avoiding read validation.


\subsection{Optimization of S-LTX and S-OCC}
\textbf{S-LTX: Read area declaration.}
The first optimization method is the read area declaration.
To alleviate the read wait for write validation, transactions declare an area that will not be read by the transaction itself at the start of the transaction (the area to be read is specified).
This eliminates the possibility of write validation conflicts in low-priority transactions, allowing them to commit without waiting for a high-priority transaction to complete, i.e., without a read wait.

\noindent
\textbf{S-LTX: On-demand version order determination.}
\quad
The second optimization method is to determine the version order on demand within the same epoch.
The version ordering of S-LTX transactions within the same epoch is dynamically determined, enabling aggressive order forwarding. When a later transaction in the serialization order has already written a version, the corresponding writes and logging of earlier transactions can be omitted by exploiting the non-visible write rule~\cite{nakazono2019invisiblewriterule}.

\begin{figure}[tb]
 \centering
 \includegraphics[scale=0.110]{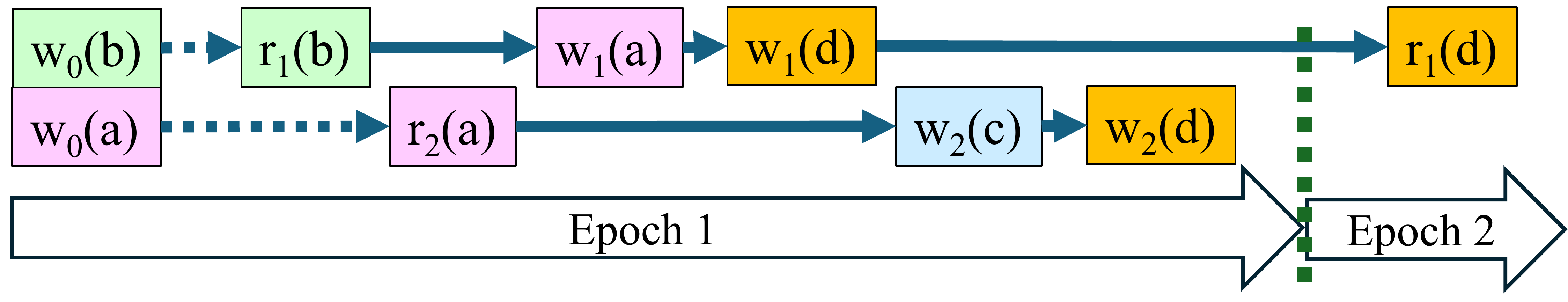}
 \caption{\small Order Forwarding of S-LTX}
 \label{fig:Running Example2}
\end{figure}

An example of this method is shown in Figure~\ref{fig:Running Example2}. Within the same epoch, $t_1$ and $t_2$ are running concurrently, where $t_1$ has already decided to commit and $t_2$ is about to make its commit decision. In this situation, $w_2(d)$ is determined to precede $w_1(d)$ in the serialization order by order forwarding, and it is guaranteed that no transaction will ever read $w_2(d)$. This is because S-OCC transactions always read at least the version written by $w_1(d)$
 or a version placed after it in serialization order, and S-LTX transactions read from a snapshot constructed on an epoch basis. Therefore, $w_2(d)$ satisfies the conditions of the non-visible write rule, and both its logging and its write to the in-memory table can be safely omitted.

\noindent
\textbf{S-LTX: Lower bound of serialization order.}
\quad
The third optimization method is to use the epoch as the lower bound of the serialization order.
Theoretically, the lower bound of the serialization order is the transaction's begin timestamp at the start of the chain of anti-dependency, but since an S-LTX transaction performs staging at the start, the begin timestamp matches the epoch information. To compress information and improve memory cache efficiency, the implementation uses epoch information as the lower bound for serialization order.

\noindent
\textbf{S-OCC: Reuse of deleted records.}
\quad
When an application executes SQL to convert a deleted record into an inserted record, multiple transactions are generated: one for the delete and one for the insert. Thus, a mechanism is required to reuse deleted records without unhooking them from the index. Shirakami also deals with deleted records as transactional reads when they are observed.

\subsection{Optimization of Write Preservation}
\textcolor{black}{
Simply employing a standard reader-writer lock is insufficient to meet our requirements. 
}
For example, when $t_1$, running on the S-LTX protocol, declares write preservation for area 1 and area 2, $t_2$, running on the S-OCC protocol, which should have a newer transaction order, observes area 1 and area 2 in this order.
In this case, we clearly need a mutual-exclusion scheme for read and write operations, similar to the reader-writer lock in 2PL.
We encounter two issues when using the reader-writer lock.

The first issue is cache misses caused by record-local lock management.
The S-OCC protocol is expected to process a large number of short transactions per unit of time.
Each time a transaction processes a record, it must read or write the specific data object within the reader-writer lock.
Even for the read-locking operation, a transaction must increase or decrease the read lock's reference count.
It may incur performance degradation~\cite{CCBench, wang2016mostly}.
The second issue is the blocking of S-LTX. If multiple S-OCC transactions continuously hold write locks, then an S-LTX transaction may not be able to acquire the lock, preventing it from initiating a transaction, which may cause starvation.

To solve these problems, our WP framework works as follows. Assume that an S-LTX transaction performs WPs at the start of the transaction (say, at epoch $N$). This is scheduled for the future, and these WPs will be effective at epoch $N+1$.
In this way, an S-OCC transaction running at epoch $N + 1$ can observe these WPs, which can be used to reduce unnecessary aborts, since the original Silo checks the lock bit in the read phase.
We assume that the number of concurrent long transactions is lower than that of concurrent short transactions in real-world use cases, such as our motivating applications.
Thus, locking and unlocking related to epoch management are unlikely to cause performance degradation due to cache pollution.

The registration, observation, and deletion of WPs are performed concurrently.
Locking and unlocking are necessary to avoid unsafe memory accesses.
As described at the beginning of this Section, WP information is handled at the table-level granularity.
Thus, it may incur performance degradation due to the locking operation.
To suppress the inefficiency, we take the following approaches.
First, we adopt an optimistic lock mechanism.
We use a 64-bit lock object, and the least significant bit (LSB) is used as the lock bit.
The remainder is used for version information.
0 indicates unlocked, and 1 indicates locked.
When modifying a WP object, a transaction waits until the lock bit is 0.
Once it observes the 0 state, it changes the lock bit from 0 to 1 atomically, modifies the WP object, sets the lock bit to 0, and increments the version at the same time.
When a transaction observes the WP object, it confirms that the lock bit is 0 before its observation.
After the observation, it immediately confirms that the lock bit is 0.
%
A transaction repeats this procedure until the check succeeds.

Second, we use fixed-length data.
We implement WP objects as arrays whose length equals the number of possible concurrent worker threads.
Each element is paired with an epoch of an S-LTX transaction to be activated and an S-LTX transaction ID.
This avoids unsafe memory accesses caused by changes in data size.

\textcolor{black}{
\subsection{Safety and Liveness}
\noindent   
\textbf{S-OCC safety.}
\quad
S-OCC is based on Silo~\cite{tu2013speedy}. Following it, S-OCC commits in conflict-serializable order via write locking and read-set validation; this property is preserved since S-OCC's modifications relative to Silo (additional WP checks against S-LTX and reader-epoch metadata) only add abort conditions.
\\
\noindent   
\textbf{S-LTX safety.}
\quad 
Each committed S-LTX is assigned a serialization epoch and an intra-epoch position determined by priority and order forwarding. This argument follows the same dependency-graph reasoning as SSN~\cite{kim2016ermia, wang2015serial}, adapted to a begin-timestamp basis: the resulting order is consistent with the read-from and anti-dependency relations observed at validation time. Shirakami's invariant that concurrent S-LTX transactions read only pre-committed versions eliminates reads-from edges among concurrent S-LTX transactions, so transitive cycles do not arise. When no consistent placement exists, the lower-priority S-LTX aborts.
\\
\noindent
\textbf{Hybrid safety.}
\quad
Within an epoch, all S-LTX transactions are ordered before all
S-OCC transactions. Cross-protocol conflicts are mediated by two mechanisms
($\S$\ref{sec:Phantom Avoidance OCC}): (i) S-OCC observes S-LTX's WP at read and commit time and aborts on conflict, ensuring S-OCC reads remain consistent with concurrent
S-LTX writes; (ii) when an S-LTX writes a record that an already-committed S-OCC has read, the S-LTX aborts via a conservative 2-sided reader-epoch check, preventing committed S-OCC reads from being invalidated.
Note that this write validation is required only when the S-LTX has performed order forwarding: a non-forwarding S-LTX never precedes any committed transaction in the serialization order—its WP, effective from its serialization epoch, forces conflicting S-OCC reads in or after that epoch to abort, and committed higher-priority S-LTX transactions are ordered before it—so its writes cannot invalidate any committed read.
Thus, every committed execution admits a serial order consistent with the assigned epochs and intra-epoch ordering.
\\
\noindent
\textbf{Liveness.}
\quad
As in Silo~\cite{tu2013speedy}, S-OCC alone does not guarantee liveness. The highest-priority S-LTX, however, is never aborted by conflicts: concurrent S-OCC transactions observe its WP and abort themselves, and lower-priority S-LTX transactions are placed by order forwarding or aborted in a way that never invalidates the highest-priority transaction's serialization position. Shirakami can therefore provide system-wide liveness by retrying starved S-OCC transactions as new S-LTX transactions and by controlling S-LTX submission priority at the application level.
}
\section{Evaluation of Tsurugi Relational Database System}
\label{sec:Evaluation with PostgreSQL}

%
\subsection{Design of Tsurugi}
\label{sec:tsurugi}
\begin{figure}[tb]
 \centering
 \includegraphics[scale=0.086]{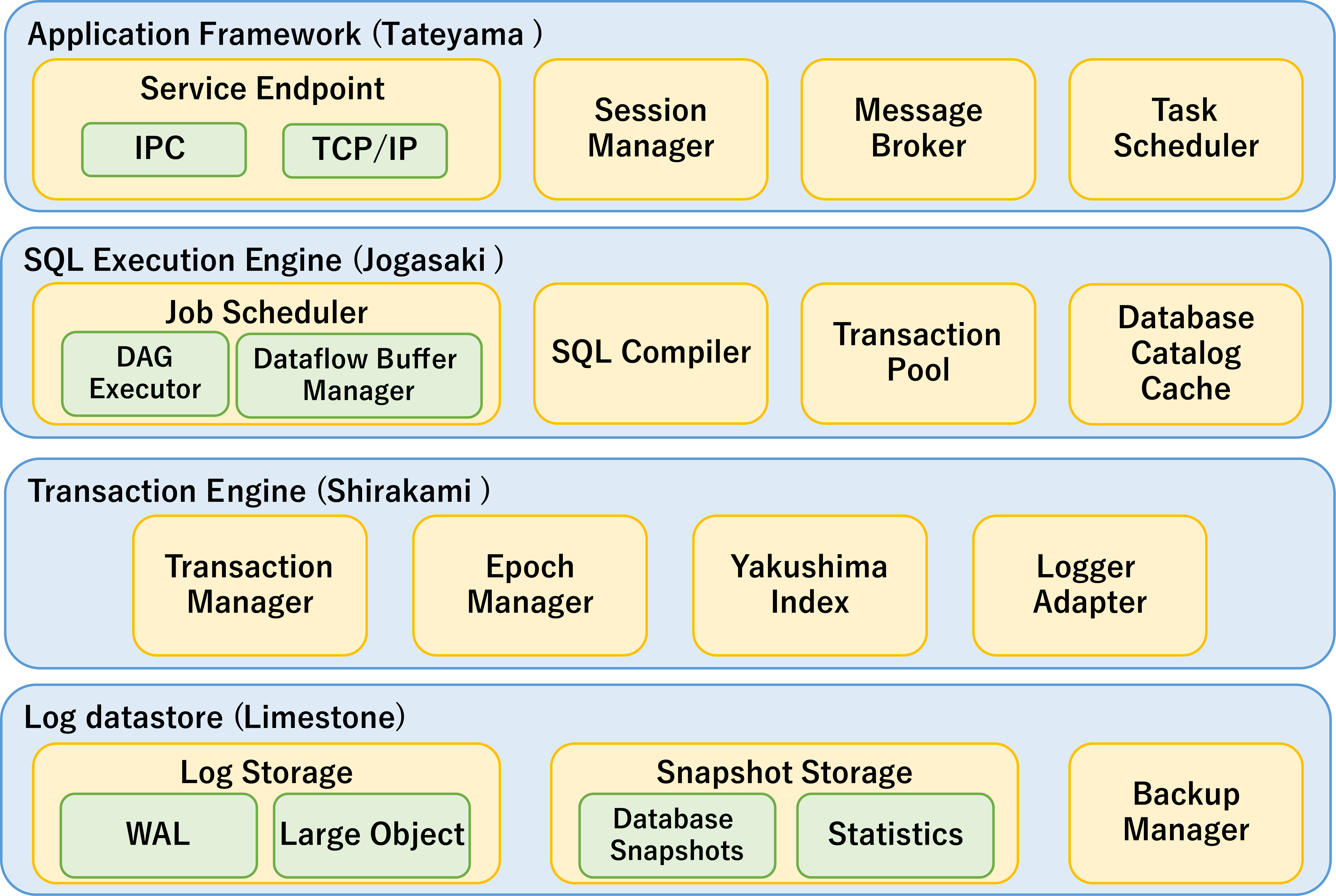}
 \caption{Tsurugi Architecture}
 \label{fig:Tsurugi Architecture}
\end{figure}

Shirakami is the transaction processing module of Tsurugi, a \textcolor{black}{relational database management system (RDBMS)} designed to run with many cores and large memory. It is built on modern architecture and is component-based, rather than monolithic.
Currently, Tsurugi is a main-memory database system, and it does not support larger-than-memory data.
Figure \ref{fig:Tsurugi Architecture} shows Tsurugi's architecture. Tateyama, the application infrastructure, provides a framework to serve various database components and loosely link them, and Jogasaki provides services to operate Tsurugi with SQL and is responsible for database operations in conjunction with the transaction engine. The transaction engine itself has a key-value store (KVS) interface and operates the database through this interface. Limestone provides services to persist the contents of transactions.
Tsurugi is available online~\cite{tsurugi_git}.

\noindent
\textbf{Yakushima: Index.}
\quad
We require an index structure to provide high performance for reads, writes, and range scans under concurrent access. To satisfy this demand, we designed and implemented Yakushima, which is based on Masstree~\cite{mao2012cache}.
The value stored in Yakushima is a pointer to a database record object, so inline optimization is used to directly embed a value object of less than 8 bytes into the value pointer.
The data layout was designed with the CPU cache architecture in mind.
In point search and range scan, optimizations were introduced to shorten the retry distance in the case of failure of optimistic verification during tree traversal. The code is available online~\cite{yakushima_git}.

\noindent
\textbf{Jogasaki: SQL Engine.}
\quad
Jogasaki provides services to operate Tsurugi with SQL and is responsible for database operations in conjunction with the transaction engine. SQL is represented by the SQL compiler as a directed acyclic graph (DAG) dataflow with various relational operators (or parts of them) at the top. The job scheduler executes SQL by processing the dataflow in parallel. The job scheduler mainly consists of a DAG executor and a dataflow buffer manager, with the former controlling the execution order of each vertex and the latter passing data between vertices.

\noindent
\textbf{Limestone: Log Datastore.}
\quad
Limestone manages log storage and snapshot storage. The former stores the contents of transaction processing and is written immediately after each transaction is completed. The latter reorganizes the contents of \textcolor{black}{write-ahead logging (WAL)~\cite{P-WAL}} to provide a snapshot suitable for reading and builds a snapshot asynchronously after writing to WAL.

\noindent
\textbf{Interface:}~The benchmark workloads in this Section are implemented using SQL.
The Tsurugi and PostgreSQL benchmark programs were written separately to account for differences in supported features and APIs.
In particular, the PostgreSQL implementation uses JDBC, whereas the Tsurugi implementation uses a Tsurugi-specific API.
Therefore, the comparison should be interpreted as a workload-level comparison under the same benchmark specification, rather than as the execution of the same program through a common interface.

\subsection{Phone Billing Benchmark}
\begin{figure}[tb]
 \centering
 \includegraphics[scale=0.21]{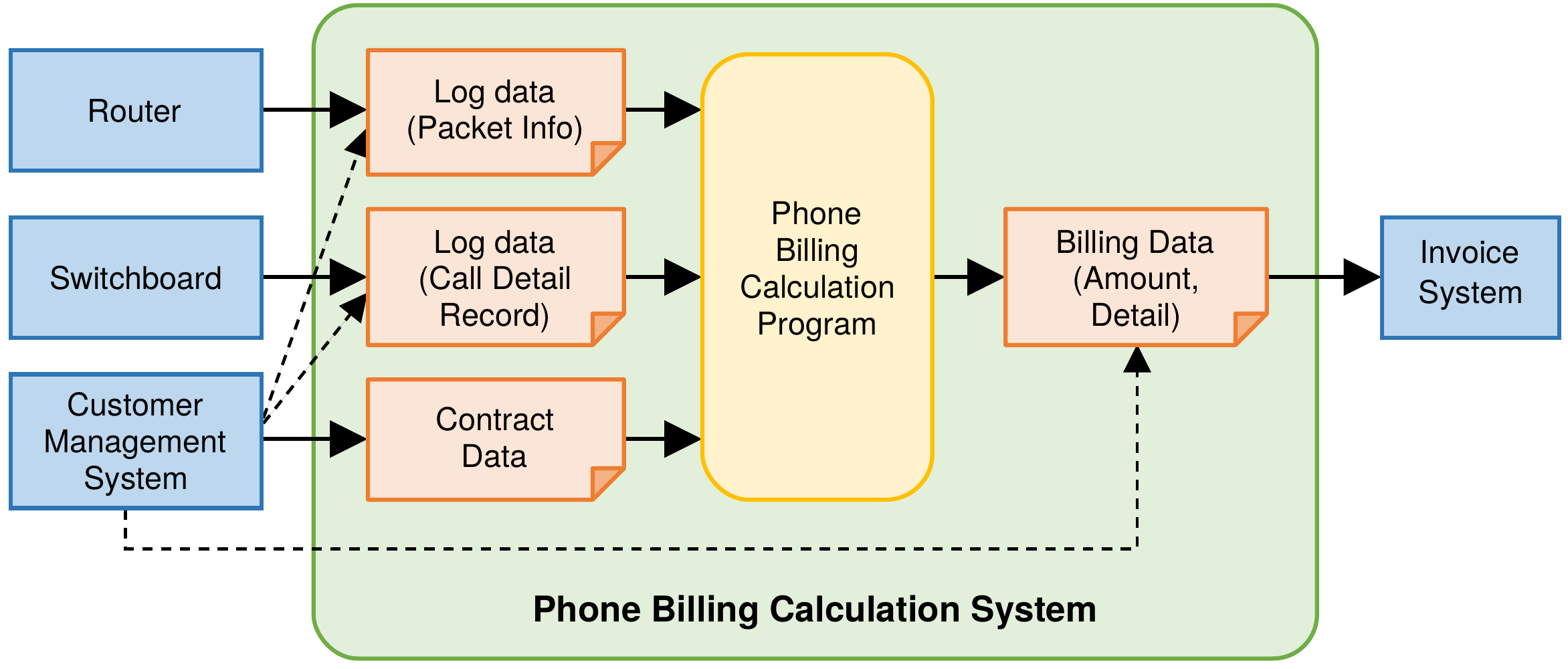}
 \caption{Phone Billing System}
 \label{fig:Billing System}
\end{figure}

We describe a benchmark for telecommunication-based phone billing. This benchmark is based on a phone billing system for a fictitious telecommunications company described in $\S$\ref{sec:Motivating Use Cases: Long Transactions}.
The architecture is illustrated in Figure~\ref{fig:Billing System}.

The telecommunications company provides cellular and fixed-line telephone services, as well as Internet access.
This system collects log information from routers, switches, and other devices. It is linked to a customer management system and maintains contract information. The system performs rate calculations in batches, by date and time, and links the calculation results to the customer management system and the billing system.
The customer management system refers to the rate calculation results when responding to user inquiries. The billing system retrieves the data necessary for billing processing from the rate calculation results and processes the billing.

The assumed business workflow is as follows. The system collects and stores log data generated by routers and switching equipment in real time. It is also linked to the customer management system. Among the customer data maintained by the customer management system, a copy of the data required for rate calculations is maintained as contract information, and additions/updates are reflected as needed.
The system also supports exceptional operations, in which log data and rate calculation results are referenced and updated upon customer management system requests.

The full business scenario also includes periodic rate calculations. A batch job is executed daily to calculate the charges for the current and previous months. The data for the current month is updated repeatedly as historical data accumulates. The data for the previous month is recalculated up to the closing date and is no longer updated thereafter.

\textcolor{black}{
Billing transactions require a SERIALIZABLE isolation level. Billing calculations should be derived from a single consistent state of contracts and history, and with READ COMMITTED, the visibility of parallel online transactions fluctuates on a statement-by-statement basis, and anomalies occur, such as new records not included in the history list read by the batch not being reflected in the billing calculation, which results in billing errors.
}

\subsubsection{Billing Calculation}
In the full target system, call and packet charges are included in the billing workflow. However, the benchmark in this paper focuses on the call-charge calculation path.
In the benchmark, the batch job first retrieves all valid contracts for the target month from the contract master. Then, each contract is processed as one transaction (i.e., one contract = one transaction). For each contract, the benchmark fetches the corresponding call history records, calculates call charges according to the contract, and writes the billing results.
This design models long-running batch updates with contract-wise processing, which is the main purpose of this benchmark.

Packet charge calculation (including matching packet logs with IP/contract information) is part of the assumed business scenario, but it is not implemented in the current benchmark.

\subsubsection{Processed Data Size}
The following values describe the assumed scale of the target telecommunication business and motivate the benchmark design. 
%
First, contract information is estimated at about 5 million records (approximately 5~GB). Next, the switching-system log data is considered.
The log contains call history records, including the caller's phone number, the callee's phone number, the call start time, the call duration, and related attributes. Each record is about 100 bytes. If 10 million records are generated per day and one year of history is retained, the total volume becomes very large.

\textcolor{black}{
Router log data is also assumed in the target business scenario. This log contains packet information, including source IP, destination IP, port, timestamp, and packet size. Each record is about 30 bytes, and 100 billion records are assumed per day. 
}
Rate calculation results are also required. Assuming 5~KB per active subscription and 5 million active subscriptions, the resulting data size is approximately 25~GB.

Note that for the benchmark used in this paper, the dataset size is much smaller than the business-scale assumption mentioned above. 
\textcolor{black}{In this paper, we measured three history record size scales, 1$\times$, 2$\times$, and 3$\times$, where 1$\times$ corresponds to 120 million history records.
Table~\ref{tab:phonebench-dataset-large} shows the dataset size for the 3$\times$ configuration. The total data size ranges from 23 GB (1$\times$) to 70 GB (3$\times$).}

\begin{table}[tb]
\centering
\caption{Dataset Size of Phone Billing Benchmark}
\label{tab:phonebench-dataset-large}
\small
\begin{tabular}{llrrll}
\hline
{\footnotesize Scale}& Table & Rows & Size (MB) & {\footnotesize CRUD} & Note \\
\hline
3$\times$ & history    & 360,000,000 & 70,310 & RU & before batch \\
   &            &(360,009,200)&(82,554)&    & after batch \\
   & contracts  &      10,000 &      1 & R  & \\
   & billing    &       5,258 &      1 & C  & after batch \\
\cline{2-6}
   & Total      & 360,015,258 & \textcolor{black}{70,312} &    & \\
\hline
\end{tabular}
\end{table}

\subsubsection{Result}
\begin{table}[tb]
\centering
\caption{Environment for RDBMS benchmark}\vskip5pt
\label{tab:machine-macro-phone}
\begin{tabular}{ll}
\hline
CPU &  INTEL(R) XEON(R) PLATINUM 8580 $\times$ 2\\
Cores & 120 physical, 240 logical\\
DRAM & 2.0TB \\
OS & Ubuntu 24.04.4 LTS \\
Kernel & 6.8.0-110-generic\\
\hline
\end{tabular}
\end{table}

\begin{figure}[tb]
 \centering
 \includegraphics[width=0.85\linewidth]{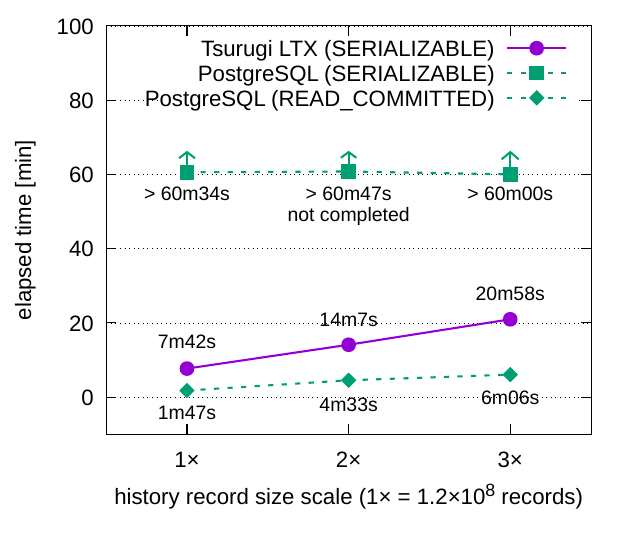}
 \caption{Result of Phone Billing Benchmark}
 \label{fig:Result of Phone Billing Benchmark}
\end{figure}
The experimental environment is shown in Table~\ref{tab:machine-macro-phone}.
Figure~\ref{fig:Result of Phone Billing Benchmark} shows the measured elapsed time of the long batch transactions in the phone billing benchmark under different \textcolor{black}{history record size scales}.
\textcolor{black}{The results indicate that Tsurugi LTX completed all runs within the time limit, and its elapsed time increased almost proportionally to the data scale.
PostgreSQL SERIALIZABLE took around 60 minutes for all scales and did not complete within the practical time limit.
At the $1\times$ scale, Tsurugi LTX was at least 7.9$\times$ faster than PostgreSQL SERIALIZABLE.
PostgreSQL READ\_COMMITTED achieved shorter elapsed times than Tsurugi LTX, but it produced anomalies, which are unacceptable.}

\subsection{\textcolor{black}{Bill of Materials Benchmark (BoMB)}}
We use a food manufacturing company that produces bread nationwide as the reference when designing the BoMB. In the target company, the supply chain can be disrupted by various factors, such as climate change, and the prices
of raw materials often change. Therefore, it is necessary to accurately estimate manufacturing costs and schedule an optimal production plan and supply chain. To reflect these requirements, the BoMB is configured to assume a system capable of on-demand inventory control, cost control, and production planning.

We assume that the target system consists of a \textcolor{black}{manufacturing resource planning (MRP)} system that manages the products and resources needed for manufacturing, and a perpetual inventory management system that continuously manages inventory. The MRP consists of cost management, budgeting, demand planning, and \textcolor{black}{supply chain management (SCM)} modules, and each module accesses one database. Cost management generates the most complicated workload among these modules because it includes
long-running update transactions. Thus, for BoMB, we focus on emulating the workload of the cost management module.

The BoMB workload has six transactions directly related to product costing and its input/output: L1 and S1, S2, S3, S4, and S5.
L and S stand for long and short, respectively. These transactions generally occur in manufacturing industries and can be widely applied beyond bakeries.
\textcolor{black}{
BoMB requires a SERIALIZABLE isolation level, and the details of this benchmark are described in \cite{oze-vldb}.
}
%

\begin{figure}[tb]
 \centering
 \includegraphics[width=1\linewidth]{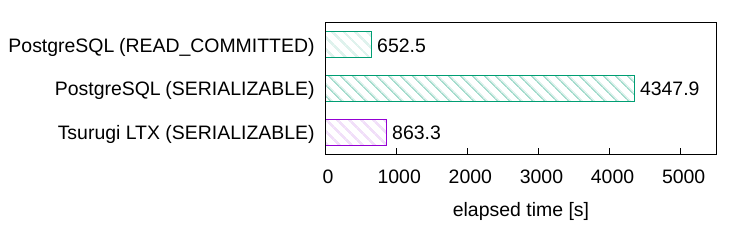}
 \caption{Result of Bill of Materials Benchmark}
 \label{fig:Result of Bill of Materials Benchmark}
\end{figure}

We evaluated the performance of Tsurugi and PostgreSQL using the BoMB benchmark.
The experimental environment is summarized in Table~\ref{tab:machine-macro-phone}.
\textcolor{black}{
The dataset size of the cost accounting benchmark is shown in Table~\ref{tab:cost-accounting-dataset-large10}.
The dataset consists of 111 million records in total and occupies 17~GB.}
The results of the experiments are shown in Figure~\ref{fig:Result of Bill of Materials Benchmark}. Tsurugi is \textcolor{black}{5.0}$\times$ faster than PostgreSQL. When we set PostgreSQL's isolation level to READ COMMITTED, PostgreSQL is faster than Tsurugi; however, it generates unacceptable anomalies.

\begin{table}[tb]
\centering
\caption{Dataset Size of Bill of Materials Benchmark}
\label{tab:cost-accounting-dataset-large10}
\small
\begin{tabular}{lrrl}
\hline
Table & Rows & Size (MB) & CRUD \\
\hline
item\_master               & 18,000,000 & 2,520 & R  \\
item\_construction\_master  & 13,757,911 & 1,837 & R  \\
item\_manufacturing\_master &  1,600,175 &   178 & R  \\
cost\_master               & 30,000,000 & 3,407 & R  \\
stock\_history             & 30,000,000 & 4,654 & -- \\
result\_table              & 17,595,413 & 4,102 & CD \\
\hline
Total                      &110,953,499 & 16,698 &    \\
\hline
\end{tabular}
\end{table}

\subsection{Results of TPC-C}
\label{sec:tpcc-results}

This Section reports TPC-C results for reference.
We used \texttt{tsubakuro\allowbreak-example}\footnote{\url{https://github.com/project-tsurugi/tsubakuro-examples}} as the benchmark driver for the Tsurugi RDBMS.
This benchmark does not invoke Shirakami directly; instead, it issues requests from outside the Tsurugi system and executes transactions by sending SQL statements.

For comparison, we also report TPC-C results on PostgreSQL.
For PostgreSQL, we used DBT-2 as the benchmark implementation.
Because Tsurugi and PostgreSQL were evaluated using different benchmark programs (i.e., \texttt{tsubakuro-example} vs.\ DBT-2) and thus not strictly under identical conditions, the absolute numbers should not be interpreted as directly comparable; they are provided only as a reference point.

\begin{figure}[tb]
 \centering
 \includegraphics[width=0.8\linewidth]{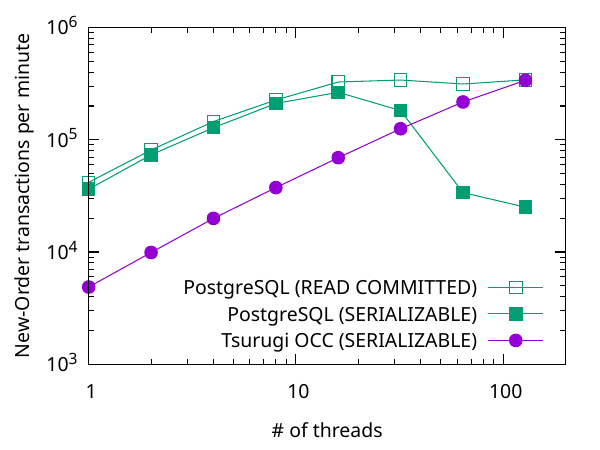}
 \caption{Result of TPC-C Benchmark}
 \label{fig:tpcc}
\end{figure}

Figure~\ref{fig:tpcc} summarizes the results.
For PostgreSQL, we evaluated two isolation levels: READ COMMITTED and SERIALIZABLE.
For Tsurugi, we ran with OCC, which provides SERIALIZABLE execution in our setting.
PostgreSQL throughput saturates at around 16 threads; under SERIALIZABLE, performance declines when the thread count exceeds 16.
In contrast, Tsurugi continues to scale up to 128 threads, with throughput increasing as the number of worker threads grows.

\section{Evaluation of Shirakami}
\label{sec:eval}

\subsection{Environment}
We implemented the Shirakami transaction processing system in C++. Our code for Shirakami is available online~\cite{shirakami-github}.
We evaluate the performance of Shirakami. This means that we bypass the SQL processing layer and interact with Shirakami through the key-value interface.
Machine specifications for the evaluation are shown in Table~\ref{tab:machine}. In the evaluation, we used 32 cores.

\begin{table}
\centering
\caption{Environment for Shirakami Transaction Engine}\vskip5pt
\label{tab:machine}
\begin{tabular}{ll}
\hline
CPU & Intel Xeon Gold 6448Y 2.10GHz \\
Cores & 32 physical, 64 logical\\
DRAM & 1.5TB \\
OS & Ubuntu 22.04.5 LTS \\
Kernel & 5.15.0-131 \\
\hline
\end{tabular}
\end{table}

\subsection{Workload}
We evaluate the performance of S-LTX, which contains both short and long transactions.
S-LTX aims to prevent long transactions from being repeatedly aborted by short transactions.
To highlight this advantage, experiments are conducted in which many short transactions and a few long transactions are executed concurrently.

Standard benchmarks such as YCSB and TPC-C do not include such mixed workloads.
To operate such a workload, we design two kinds of YCSB-based workloads.
We call them workload-1 and workload-2, respectively.
Workload-1 (WL1) includes short transactions and long transactions.
For each short transaction, there are 10 operations, with a read-write ratio of 50\%:50\%.
For each long transaction, its read-write ratio is 50\%:50\%, and we varied the number of operations in a transaction:
125, 250, 500, 1000, 2000, 4000, and 8000 operations.
Workload-2 (WL2) includes short transactions and long transactions.
For each short transaction, there are 10 operations, all of which are insert operations.
For each long transaction, there are only scan operations, and we varied the range of scan operations in a transaction:
100, 1000, 2500, 5000, 10000, 20000, 40000, and 100000.

The long transaction ratio, which refers to the fraction of long transactions, is set to 0.01, 0.001, and 0.0001.
Each experiment compared LTX and OCC for long TX execution.
The database initially contained 100,000 records with uniform access (skew = 0).
All experiments ran for 30~s.

\begin{figure*}[th]
 \centering
 \begin{minipage}[t]{1\linewidth}
   \centering
   {\small Long TX = OCC}
   \begin{minipage}[c]{0.25\linewidth}
    \includegraphics[trim=10.5cm 6.5cm 0 2.3cm,clip,width=3cm]{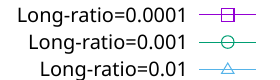}
   \end{minipage}
   {\small Long TX = LTX}
   \begin{minipage}[c]{0.25\linewidth}
    \includegraphics[trim=10.5cm 5cm 0 3.7cm,clip,width=3cm]{ycsb1-ltx-rw-w50-s0.0.pdf}
   \end{minipage}
  \caption{Legend for Figs.~\ref{fig:RW-LTX}-\ref{fig:SI-ratio}}
  \label{fig:legend}
 \end{minipage}
\\\vspace{1em}
 \begin{minipage}[b]{0.33\linewidth}
  \centering
  \includegraphics[trim=0 0 4.5cm 1.8cm,clip,width=1.0\linewidth]{ycsb1-ltx-rw-w50-s0.0.pdf}
  \caption{LongTX throughput (WL1)}
  \label{fig:RW-LTX}
 \end{minipage}
 \begin{minipage}[b]{0.33\linewidth}
  \centering
  \includegraphics[trim=0 0 4.5cm 1.8cm,clip,width=1.0\linewidth]{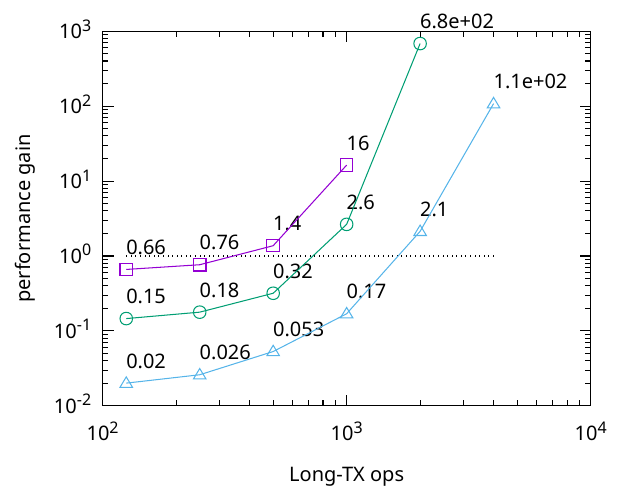}
  \caption{Performance Gain (WL1)}
  \label{fig:RW-gain}
 \end{minipage}
 \begin{minipage}[b]{0.33\linewidth}
  \centering
  \includegraphics[trim=0 0 4.5cm 1.8cm,clip,width=1.0\linewidth]{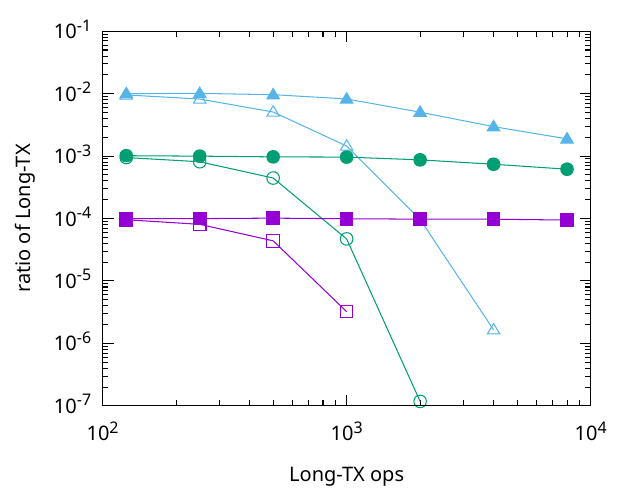}
  \caption{LongTX ratio (WL1)}
  \label{fig:RW-ratio}
 \end{minipage}
\\\vspace{1em}
 \begin{minipage}[b]{0.33\linewidth}
  \centering
  \includegraphics[trim=0 0 4.5cm 1.8cm,clip,width=1.0\linewidth]{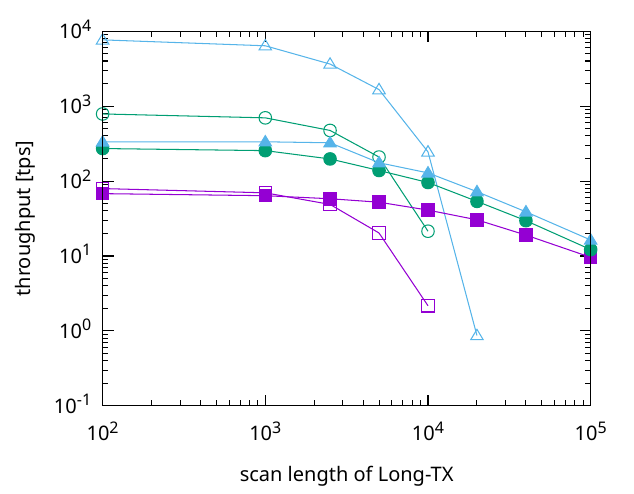}
  \caption{LongTX throughput (WL2)}
  \label{fig:SI-LTX}
 \end{minipage}
 \begin{minipage}[b]{0.33\linewidth}
  \centering
  \includegraphics[trim=0 0 4.5cm 1.8cm,clip,width=1.0\linewidth]{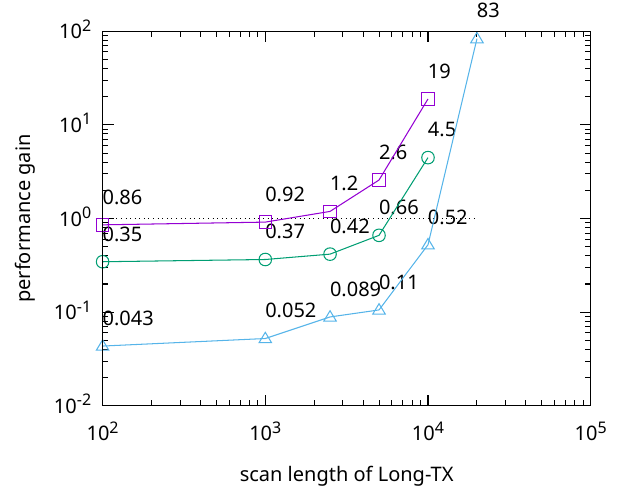}
  \caption{Performance Gain (WL2)}
  \label{fig:SI-gain}
 \end{minipage}
 \begin{minipage}[b]{0.33\linewidth}
  \centering
  \includegraphics[trim=0 0 4.5cm 1.8cm,clip,width=1.0\linewidth]{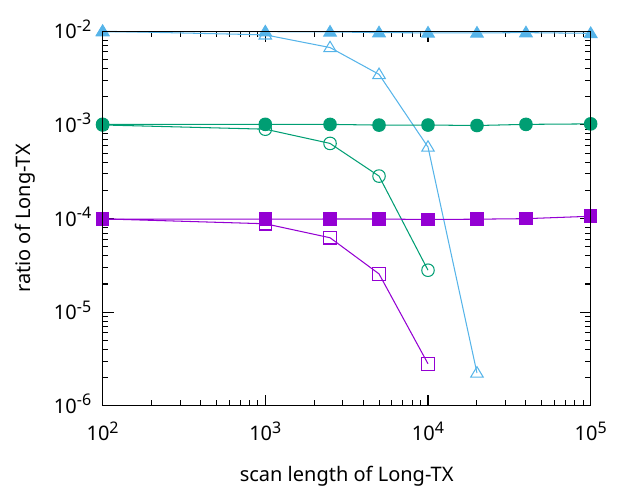}
  \caption{LongTX ratio (WL2)}
  \label{fig:SI-ratio}
 \end{minipage}
\end{figure*}

Figures~\ref{fig:RW-LTX}--\ref{fig:SI-ratio} summarize the results.
Figures~\ref{fig:RW-LTX}--\ref{fig:RW-ratio} show results for WL1 (Read-Write (Update) 50\%).
Figures~\ref{fig:SI-LTX}--\ref{fig:SI-ratio} show results for WL2 (Insert-Scan).
Each workload reports:
(1) Long-TX throughput (LTX vs. OCC) (Figures~\ref{fig:RW-LTX},\ref{fig:SI-LTX}),
(2) Relative throughput (LTX/OCC) (Figures~\ref{fig:RW-gain},\ref{fig:SI-gain}),
and (3) Commit ratio of long TXs to total TXs (Figures~\ref{fig:RW-ratio},\ref{fig:SI-ratio}).

\noindent
\textbf{Results on Workload 1 (WL1).}
\quad
Under Long-ratio = 0.0001, Figure~\ref{fig:RW-LTX} shows that OCC throughput drops sharply when the long TX length increases from 500 to 1000 ops, eventually reaching zero.
LTX maintains a nonzero throughput up to 8000 ops, indicating successful completion of long TXs even under contention-intensive conditions.
As shown in Figure~\ref{fig:RW-gain}, the performance improvement (LTX/OCC) reaches 16$\times$ at 1000 ops. 
The gain grows up to 680$\times$ because the baseline goes to nearly zero.
Beyond this size, OCC fails to commit any long TXs.
Figure~\ref{fig:RW-ratio} demonstrates that OCC’s long-TX commit ratio decreases rapidly with TX length, whereas LTX sustains the configured ratio of 0.0001.
For Long-ratio = 0.001 and 0.01, the crossover point at which LTX overtakes OCC shifts toward larger TX sizes.
For smaller long TXs ($< 1000$ ops), LTX incurs overhead (LTX/OCC = 0.32--0.02).
Thus, LTX is most effective when long TXs are large and infrequent.

\noindent
\textbf{Results on Workload 2 (WL2).}
\quad
Figures~\ref{fig:SI-LTX}--\ref{fig:SI-ratio} (Short Insert-Long Scan) exhibit trends consistent with WL1.
When scan ranges are large and long TXs are rare, OCC fails to commit due to conflicts, while LTX maintains stable throughput.
These results confirm that LTX effectively supports mixed workloads, including heavy, low-frequency long TXs and high-frequency, short TXs.

Overall, Shirakami-LTX significantly improves the success rate and throughput of long transactions in mixed workloads.
While S-LTX introduces modest overhead when long transactions are frequent or short, it achieves strong progress \textcolor{black}{guarantees} and substantial performance gains.

\section{Related Work}
\label{sec:related}

\subsection{Concurrency Control}
\subsubsection{Protocols for Short Transactions}~
A wide array of concurrency control protocols has been proposed to push the boundaries of transaction processing
~\cite{eswaran1976notions, yu2016tictoc, STOv2, wang2015serial, wang2016mostly, Chen22plor, epic, vandevoort24, tianzheng25}.
A defining characteristic of these studies is their reliance on speculative execution, managing dependencies to minimize blocking.
While these protocols have demonstrated unprecedented performance on standard benchmarks such as YCSB and TPC-C, the workloads they run primarily consist of short transactions. Consequently, these speculative methods often fail to maintain high throughput when transaction sizes are large or execution times are prolonged.
To address these inherent weaknesses, deterministic approaches~\cite{Thomson12calvin, Lu2020Aria, Fan19oceanvista} have emerged as a robust alternative. By pre-declaring read/write sets beforehand, these systems can successfully handle large transactions without the penalty of frequent aborts.
These protocols do not address long read-write transactions.

\noindent
\subsubsection{Protocols for Long Transactions}
Recent advancements have introduced specialized mechanisms to bridge the gap between short and long transactions. Oze~\cite{oze-vldb} explores the MVSR, which offers a broader scheduling space than traditional CSR.
In the domain of graph data management, TuskFlow \cite{tuskflow} specifically targets the challenges posed by long-running read-write transactions. Its technical foundation rests on deterministic concurrency control \cite{Lu2020Aria}, which eliminates the overhead of runtime coordination. While TuskFlow demonstrates high throughput on real-world graph database benchmarks, it remains constrained by strict dataset prerequisites inherent to its deterministic protocol. Furthermore, its scheduling space is confined to CSR, potentially limiting its concurrency in highly contended environments.
An alternative approach for graph data is provided by DDI~\cite{ddi}, which adopts a different strategy by selectively utilizing multiple weak isolation levels (e.g., Read Committed and Snapshot Isolation). While this enables the execution of long read-write transactions, it meticulously ensures overall serializability through a sophisticated coordination layer.
In contrast to these specialized systems, Shirakami introduces a novel hybrid architecture that combines S-LTX and S-OCC, enabling it to deliver superior performance for both short-term and long-lived transactions.
\textcolor{black}{
The Wait-Hit protocol~\cite{Waudby} achieves high performance by circumventing complex cycle detection in favor of lightweight verification based on compressed conflict information, a technique that should prove efficient for Mammoth transactions~\cite{Mammoth}. However, while Wait-Hit’s scheduling space is restricted to CSR, Shirakami offers a broader scheduling space that extends beyond MVTO's.
}

\subsection{Modern Database Engine}
To complement the evolution of individual concurrency control protocols, significant research has focused on the architectural design of complete transaction processing systems. These systems aim to integrate theoretical advancements into production-ready environments; yet, they often inherit the specific trade-offs of their underlying mechanisms.
HiEngine~\cite{hiengine}, YugaByteDB~\cite{Yugabyte}, and LineairDB~\cite{lineairdb_git} represent sophisticated systems predominantly engineered around Optimistic Concurrency Control (OCC) or its variants. While these architectures excel in high-concurrency environments with low contention and short-lived workloads, they frequently degrade in performance—such as high abort rates or excessive memory pressure—when subjected to the resource-intensive nature of long-running transactions.

To the best of our knowledge, Tsurugi stands as the only production-grade system designed to overcome this dichotomy. 
Shirakami ensures that transient operations are not penalized by heavyweight background tasks.

\section{Conclusions}
\label{sec:concl}
Modern real-world transactional workloads, such as Bill of Materials and telecommunications Phone Billing, need to process both short and long transactions simultaneously.
Conventional concurrency control protocols do not adequately address such workloads. This is because they assume only classical workloads (i.e., YCSB and TPC-C) that contain relatively short transactions, or they consider only a specific type of workload (i.e., graph analytics~\cite{tuskflow, ddi}).

We proposed a new concurrency control protocol, Shirakami. Shirakami has two sub-protocols.
The first is the Shirakami-LTX protocol, designed for long transactions and based on multiversion concurrency control.
The second is the Shirakami-OCC protocol, designed for short transactions and based on the Silo protocol.
Shirakami naturally integrates them with the write-preservation method and epoch-based synchronization.
Shirakami is designed and implemented as a transaction-processing module in Tsurugi, a relational database management system~\cite {tsurugi_git}.

\textcolor{black}{
We evaluated Tsurugi LTX and confirmed that it outperforms PostgreSQL SERIALIZABLE, achieving at least a 7.9$\times$ speedup in the phone billing benchmark and a 5.0$\times$ speedup in the bill-of-materials benchmark.}
We also evaluated Shirakami using a YCSB benchmark with long transactions.
We compared S-LTX, a protocol tailored for long transactions, with S-OCC.
The experimental results demonstrated that S-LTX exhibited 680 times higher throughput than S-OCC.

\textcolor{black}{
In future work, we will design and implement a buffer management system to handle data that exceeds memory capacity.
As Tsurugi evolves to accommodate datasets exceeding physical memory capacity through buffer management, I/O operations will become a primary factor in extending transaction duration. We anticipate that S-LTX will demonstrate superior utility in these contexts by effectively managing the increased execution latency.
}

Tsurugi is a production system and has real use cases~\cite{formula}.
We hope it will support more use cases that require handling long read-write transactions.

\section*{Acknowledgments}
%
%
This paper is based on results obtained from a project, JPNP16007, commissioned by the New Energy and Industrial Technology Development Organization (NEDO), and from JSPS KAKENHI Grant Number 25H00446, and from JST CREST Grant Number JPMJCR24R4, and from SECOM Science and Technology Foundation.

\bibliographystyle{ACM-Reference-Format}
\bibliography{9ref}
\end{document}